# Ternary Fingerprints with Reference Odor for Fluctuation-Enhanced Sensing

Xiaoyu Yu [1,+], Laszlo B. Kish [2], Jean-Luc Seguin [3] and Maria D. King [1]

[1] Department of Biological and Agricultural Engineering, Texas A&M University, College Station, TX 77843-2117, USA.

[2] Department of Electrical and Computer Engineering, Texas A&M University, College Station, TX 77843-3128, USA.

[3] Aix Marseille Univ, University of Toulon, CNRS, IM2NP, Marseille, 13397, France.

Correspondence should be addressed to Maria D. King; mdking@tamu.edu

## Abstract

An improved method for Fluctuation Enhanced Sensing (FES) is introduced. We enhanced the old binary fingerprinting method, where the fingerprint bit values were ± 1, by introducing ternary fingerprints utilizing a reference odor. In the ternary method, the fingerprint bit values are -1, 0, and +1 where the 0 value stands for the situation where the slope of the spectrum is identical to that of the reference odor. The application of the reference odor spectrum makes the fingerprint relative to the reference. This feature increases the information entropy of the fingerprints. The method is briefly illustrated by sensing bacterial odor in cow manure isolates.

## Introduction, Fluctuation-Enhanced Sensing (FES)

Fluctuation-Enhanced Sensing (FES) [1-36] utilizes the statistical properties of microscopic random fluctuations superimposed on the classical sensor signal to generate patterns that identify the chemical composition of odors. In most of cases, the power density spectrum (PDS) of the amplified and measured fluctuations is used as identification pattern, see Figure 1.

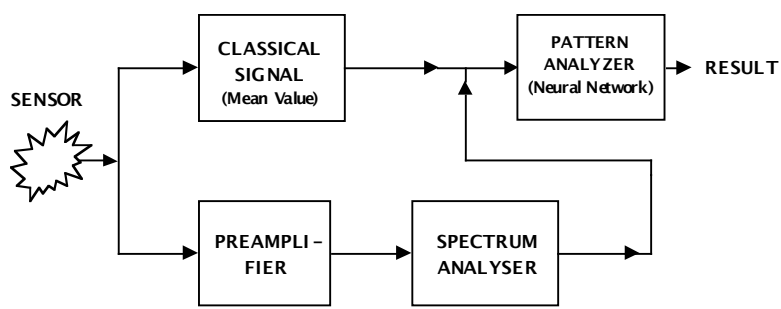

Figure 1. Fluctuation-enhanced sensing scheme with PDS

+ On leave from Sun Yat-sen University, School of Atmospheric Sciences, China.

In the case of Gaussian processes, the PDS contains all the achievable statistical information about the fluctuations. However, in the non-Gaussian case, Higher-Order Statistical (HOS) tools such as the bispectra [6,11] offer additional information.

## Materials and Methods

### Binary fingerprints

In principle, the PDS, and the bispectra (or other HOS tools) can directly be fed into a qualifier, such as a neural network, to identify the chemical composition related to this pattern. However, spectra may contain excess amounts of irrelevant information and the training of neural networks is a tremendous task as it requires executing a great number of measurements with a large variety of chemical compositions.

Thus, much simpler and more direct approaches have also been tested with good results, for example the *binary fingerprint* method that extracts a bit string from the measured PDS [23], see Figure 2. The average slope of the spectrum plotted with a log-log scale is determined by connecting the beginning and the end of the ("meaningful" part of the) PDS. Next, the same frequency band is divided into sub-bands to determine the related binary bit values. Then the local slope over these sub-bands is determined in the same way as described above.

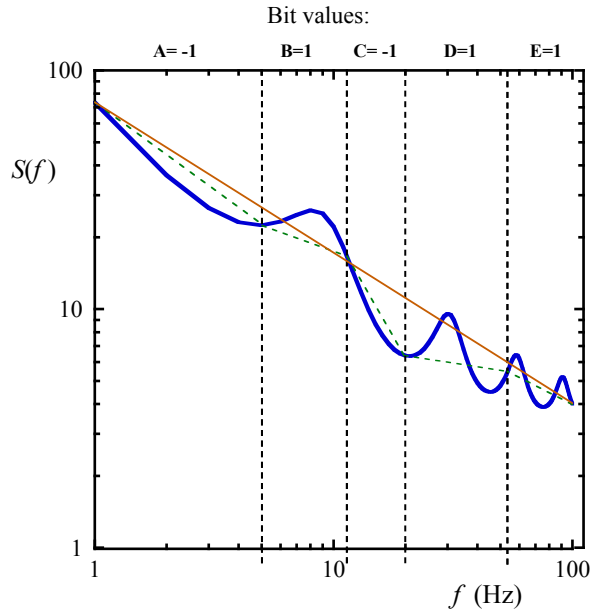

Figure 2. Generating the binary fingerprints [23]. The local slope is compared with the global slope. The bit values are -1 (bits A, C) and 1 (bits B, D, E) depending on the result of this comparison.

When the local slope is below the average, the bit value is -1, otherwise it is 1, see Figure 3.

In the present paper, we generalize the binary method to the ternary one. The new method offers additional information and ways to use comparative features with reference odors. After summarizing the previous method and introducing the new one, we demonstrate and compare them by generating these fingerprints with cow-manure related odor.

**The new method: Ternary fingerprints**

For the trinary fingerprints, we also included a reference agent. The spectrum of the reference agent serves as the reference PDS. Next, the frequency band is divided into sub-bands (similarly to the case of the binary fingerprints) to determine the related ternary bit values. Then the local slope on these sub-bands is determined in the same way as described above.

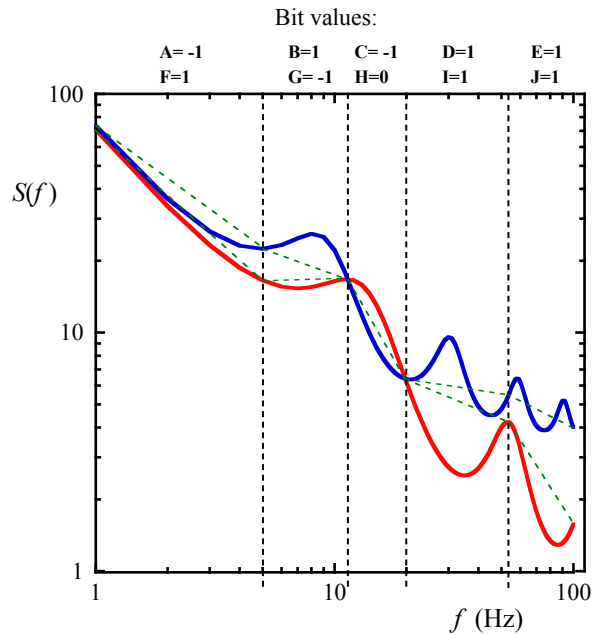

Figure 3. Generating the ternary fingerprints of the same PDS. The reference spectrum is shown with red line. When the local slope of the spectrum is greater than the local slope of the reference spectrum, the bit value is 1 (bits F, I, J). In the opposite case it is -1 (bit G). When the slopes are equal, the bit value is zero (bit H). The binary fingerprints of the PDS are also shown (bits A - E).

When the local slope with the agent is less than the local slope with the reference agent, the bit value is -1; when it is greater than the reference slope, the bit value is 1; and when the slopes are equal (this happens with a small probability depending on the resolution) the bit value is zero, shown in Figures 3 and 4.

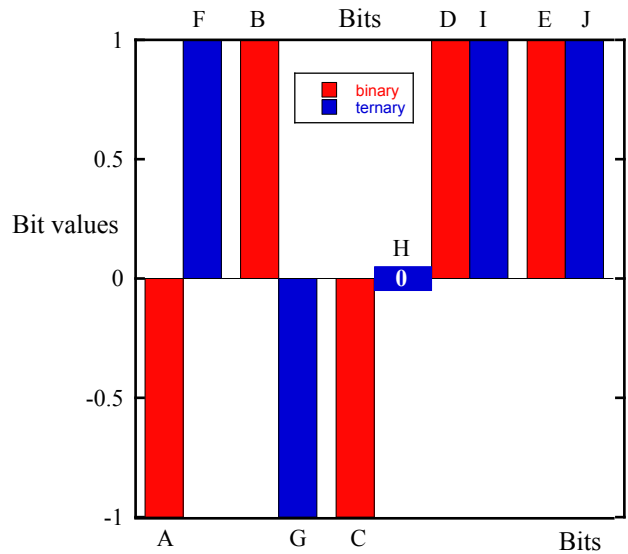

Figure 4. Joint plot of the binary and ternary fingerprints based on the source spectra shown in Figures 2 and 3.

**Demonstration with bacterial isolates from cow manure**

For the demonstration of the ternary fingerprinting method introduced above we used a bacterial strain isolated from cow manure. The Petri plates of 58 cm$^2$ with the bacteria colonies were placed in a 300 cm$^3$ sensor chamber with the sensors attached.

The microorganism used in this study was a Gram-positive toxin producing, facultative anaerobic bacteria, *Bacillus cereus*. Mid-log phase (OD600 = 0.5, optical density at 600 nm) cultures of *Bacillus cereus*, isolated from cow manure at a dairy center in Stephenville, Texas were grown in Luria Bertani (LB) medium [1] for 4 h at 37 °C and at 150 rpm. The most abundant bacterium was isolated from different manure samples and identified as *Bacillus cereus* by whole genome sequencing at TIGSS (Texas A&M University Institute for Genome Sequencing and Society). One hundred microliters of the *B. cereus* culture were spread on Difco tryptic soy agar (TSA) plates (Becton Dickinson Co., Sparks, MD), and the plates were incubated overnight at 37 °C [37]. As reference, sterile TSA plates without bacteria were also prepared. As the TSA medium itself has a strong smell, identical amounts (27 mL) of TSA medium were poured into each plastic Petri plate (VWR, Bridgeport, NJ) to maintain a constant level of background odor [37].

The metal oxide (Taguchi) sensor was a 50 nanometer thick $SnO_2$ film, sputtered on a 4x4μm² microhotplate with platinum heater and sensing electrodes. Details are described elsewhere [38].

## Results and Discussion

Figure 5 shows the measured PDS with the bacterial sample and the reference PDS which was measured using the same sensor in laboratory air. Figure 6 shows the binary and ternary fingerprints extracted from the spectra shown in Figure 5, respectively.

The reproducibility of the odor sensing systems is of great importance. Formerly binary spectra showed a good reproducibility with *Escherichia coli* and *Bacillus anthracis* (anthrax) bacterial samples [23]. Similarly, the reproducibility of the binary fingerprints for the manure isolate *Bacillus cereus* was satisfactory (Figure 7). Therefore, with the ternary fingerprint method a good reproducibility was also expected.

We tested the reproducibility of the ternary fingerprints with our sensor system and bacteria and reference samples. Figure 8 indicates that the reproducibility results of the ternary fingerprints are satisfactory.

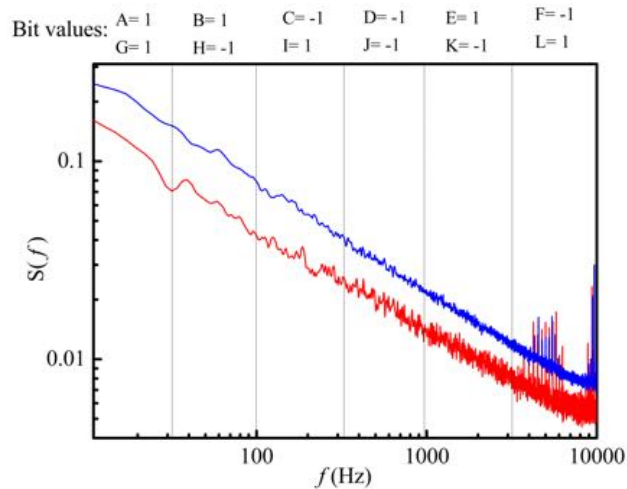

Figure 5. The PDS of the bacterium isolate measured by the Taguchi sensor [30-37] shown by the blue line. For reference spectrum the PDS was measured in laboratory air, shown by the red line.

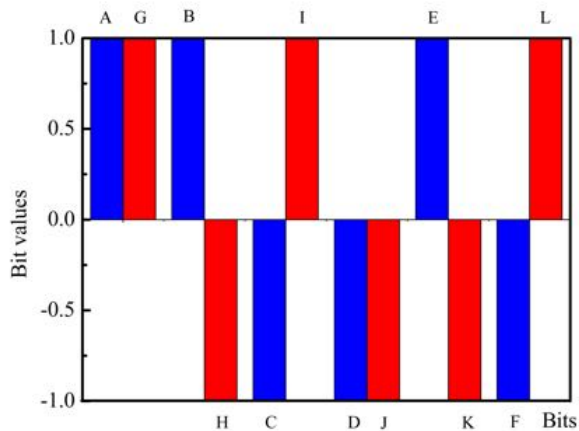

Figure 6. Binary and ternary fingerprints of the bacterium sample.

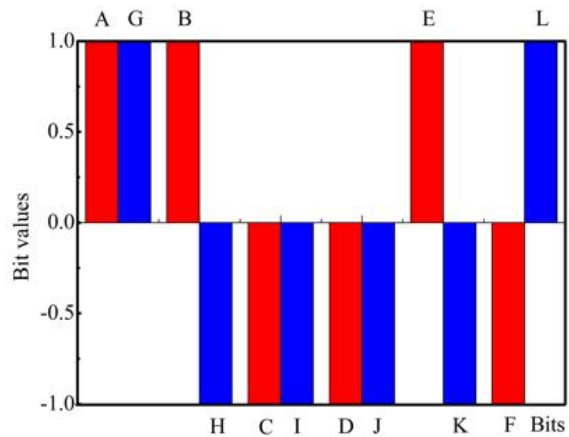

Figure 7. Reproducibility of the binary fingerprints. Red: measurement-1; Blue: measurement-2 (of the same sample).

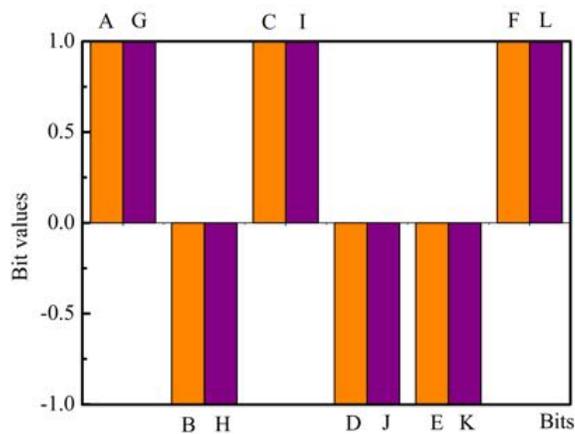

Figure 8. Reproducibility of the ternary fingerprints. Orange: measurement-1; Maroon: measurement-2 (of the same sample).

## Conclusions

An improved method for evaluating Fluctuation Enhanced Sensing (FES) results was introduced. In the ternary method, the fingerprint bit values are -1, 0, and +1 where the 0 value stands for the situation where the slope of the spectrum is identical to that of the reference odor. The application of the reference odor spectrum makes the fingerprint relative to the reference. This feature increases the information entropy the fingerprints. Measuring bacterial odor in cow manure isolates indicates good reproducibility.

## Data Availability

The data that support the findings of this study are available on request from the corresponding author, MDK.

# Conflicts of Interest

The authors declare that there is no conflict of interest regarding the publication of this paper.

# Funding Statement

XY's visit was supported by the China Research Foundation. MDK is supported by the USDA National Institute of Food and Agriculture, Hatch project TEX09746 and by the T3 grant at Texas A&M University (2019-2020). LBK is supported by the T3 grant at Texas A&M University (2019-2020).

# Acknowledgements

The authors wish to thank Tomas Fiorido for technical support in sensors fabrication.